\documentclass[aps,preprintnumbers,showpacs,twocolumn,superscriptaddress,floatfix]{revtex4}

\usepackage{amsmath}
\usepackage{latexsym}
\usepackage{graphicx}
\usepackage{epstopdf}
\usepackage{times}
\input{epsf.sty}

\newcommand{\bea}{\begin{eqnarray}}
\newcommand{\eea}{\end{eqnarray}}
\newcommand{\beq}{\begin{equation}}
\newcommand{\eeq}{\end{equation}}

\begin{document}

\title{Getting a kick from equal-mass binary black hole mergers}

\author{Michael~Koppitz}
\affiliation{
  Max-Planck-Institut f\"ur Gravitationsphysik,
  Albert-Einstein-Institut,
  Potsdam-Golm, Germany
}

\author{Denis~Pollney}
\affiliation{
  Max-Planck-Institut f\"ur Gravitationsphysik,
  Albert-Einstein-Institut,
  Potsdam-Golm, Germany
}

\author{Christian~Reisswig}
\affiliation{
  Max-Planck-Institut f\"ur Gravitationsphysik,
  Albert-Einstein-Institut,
  Potsdam-Golm, Germany
}

\author{Luciano~Rezzolla}
\affiliation{
  Max-Planck-Institut f\"ur Gravitationsphysik,
  Albert-Einstein-Institut,
  Potsdam-Golm, Germany
}
\affiliation{
  Department of Physics and Astronomy,
  Louisiana State University,
  Baton~Rouge, LA, USA
}

\author{Jonathan Thornburg}
\affiliation{
  Max-Planck-Institut f\"ur Gravitationsphysik,
  Albert-Einstein-Institut,
  Potsdam-Golm, Germany
}

\author{Peter~Diener}
\affiliation{
  Center for Computation \& Technology,
  Louisiana State University,
  Baton~Rouge, LA, USA
}
\affiliation{
  Department of Physics and Astronomy,
  Louisiana State University,
  Baton~Rouge, LA, USA
}

\author{Erik~Schnetter}
\affiliation{
  Center for Computation \& Technology,
  Louisiana State University,
  Baton~Rouge, LA, USA
}

\date{\today}


\begin{abstract}
   The final evolution of a binary black-hole system gives rise to a
   recoil velocity if an asymmetry is present in the emitted
   gravitational radiation. Measurements of this effect for
   non-spinning binaries with unequal masses have pointed out that
   kick velocities $\sim~175$ km/s can be reached for a mass ratio
   $\simeq 0.36$. However, a larger recoil can be obtained for
   equal-mass binaries if the asymmetry is provided by the
   spins. Using two independent methods we show that the merger of
   such binaries yields velocities as large as $\sim 440$ km/s for
   black holes having unequal spins that are antialigned and parallel
   to the orbital angular momentum.
\end{abstract}

\maketitle


\noindent\emph{Introduction.~} Binary black hole systems are expected
to be one of the strongest sources of gravitational waves and are
therefore the subject of intense and careful investigations. With
earth-based gravitational-wave detectors now working at
design-sensitivity and a space-borne detector in its formulation
phase, the need for reliable templates to be used in matched filtering
techniques has inspired a renewed enthusiasm in numerical
analysis. Using numerical methods developed
recently~\cite{Pretorius:2005gq, Campanelli:2005dd, Baker:2005vv}
there has been an explosion of results (see, \textit{e.g.,}
refs.~\cite{Pretorius:2006tp, Baker:2006yw, Campanelli:2006gf,
  Baker:2006ls, Baker:2006nr, Baiotti06, Diener-etal-2006a,
  Gonzales06tr, Campanelli:2006uy, Campanelli:2006vp,
  Campanelli:2006fy, Baker-etal:2007a, Thornburg-etal-2007a}).

These developments are important for at least three different reasons.
First, they allow for improved templates to be used in the analysis of
the data coming from the detectors. Second, they allow probes of
General Relativity in regimes that have previously been
inaccessible. Last but not least, they can provide, even through the
solution of the Einstein equations in vacuum, important astrophysical
information.

Together with energy and angular momentum, gravitational radiation
also carries away linear momentum. In the case of a binary system of
non-spinning black holes, a physical intuition of this loss of linear
momentum can be built rather easily. As the two bodies orbit around
the common center of mass, each will emit radiation which is
forward-beamed. Unless the two black holes have exactly the same mass,
their motion will be different, with the smaller black hole moving
more rapidly and, hence, being more efficient in beaming its
emission. The net momentum gained over an orbit is negligible if the
orbit is almost circular (the momentum loss in any direction is
essentially balanced by an equal loss in the diametrically opposite
direction), but it can become large when integrated over many orbits,
leading to a recoil that is a fraction ($\lesssim 10^{-2}$) of the
speed of light during the last portion of the orbit prior to the
merger.

A number of PN/perturbative analyses (see,
\textit{e.g.,}~\cite{Favata:2004wz, Damour-Gopakumar-2006}) have
provided estimates of this recoil velocity, while numerical-relativity
simulations~\cite{Baker:2006nr,Gonzales06tr} have recently measured it
to rather high precision, predicting a maximal kick of $175$ km/s for
a binary system of nonspinning black holes with a mass ratio $q \equiv
M_1/M_2 \simeq 0.36$, where $M_1$ and $M_2$ are the masses of the two
black holes. Such a recoil has indeed quite important astrophysical
consequences, since it could, provided it is large enough, kick the
binary out of its host environment. Clearly, a replaced or an even
missing central black hole would have dramatic consequences for the
further development of the host. Determining accurately what are the
expected escape velocities for the most typical environments hosting a
binary black hole system is rather difficult, but the estimates made
in refs.~\cite{Merritt:2002hc}, for instance, predict that the escape
velocities for dwarf galaxies and globular clusters are $\lesssim 100$
km/s, but for giant galaxies these can be $\sim 1000$ km/s.

When adopting a purely geometrical viewpoint, it is obvious that a
kick velocity should be expected in any binary system which is not
perfectly symmetric. A difference in the masses is a simple way of
producing such an asymmetry but surely not the only one. Indeed, even
an equal-mass system can be made asymmetric if the two black holes
have unequal spins. Also in this case, a simple physical intuition can
be constructed. Consider, for simplicity an equal-mass binary in which
only one member is spinning parallel to the orbital angular
momentum. As a result of the spin-induced frame dragging, the speed of
the nonspinning body will be increased and its radiation further
beamed. Using PN theory at the 2.5 order, Kidder~\cite{Kidder:1995zr}
has treated this spin-orbit interaction concluding that in the case of
a circular, non-precessing orbit, the total kick for a binary system
of arbitrary mass and spin ratio can be expressed
as~\cite{Favata:2004wz}
\beq 
\label{eq:fhh} 
|v|_{\rm kick}=c_1 \frac{q^2(1-q)}{(1+q)^5} + 
	c_2 \frac{a_2 q^2(1 - q  a_1/a_2)}{(1+q)^5}\;, 
\eeq
where $a_{1,2} \equiv S_{1,2}/M^2_{1,2}$ are the dimensionless spins
of the two black holes and these are aligned with the total orbital
angular momentum, \textit{i.e.,} ${\boldsymbol
S}_{1,2}=a_{1,2}M^2_{1,2} {\boldsymbol e}_z$ for an orbital motion in
the $(x,y)$ plane. Here, $c_{1}$ and $c_{2}$ are factors depending on
the total mass of the system and on the orbital separation at which
the system stops radiating.  This radius is difficult to determine
precisely as it lies in a region where the PN approximation is not
very accurate and is, in practice, not even a constant but, rather,
depends on both the mass and the spin ratio. Assuming for simplicity
$c_1 \simeq c_2$, expression (\ref{eq:fhh}) reveals that a substantial
contribution to the recoil velocity comes from the spins alone. In
addition, for any given $q$, it predicts a linear growth of the recoil
velocity with increasing difference in spins, yielding a kick which is
comparable with the one coming from the asymmetry in the mass. Stated
differently, when it comes to recoil velocities, the spin
contributions may be the dominant ones.


\noindent\emph{Techniques.~} The numerical evolutions have been
carried out using a conformal-traceless formulation of the Einstein
equations as described in~\cite{Alcubierre02a}, with ``$1{+}\log$''
slicing and $\Gamma$-driver shift, and advection terms applied to the
gauge conditions as suggested in~\cite{Baker:2005vv}. Spatial
differentiation is performed via straightforward finite differencing
using fourth-order stencils.  Individual apparent horizons are located
every few timesteps during the
evolution~\cite{Thornburg2003:AH-finding_nourl}. Vertex-centered
adeptive mesh-refinement (AMR) is employed using nested
grids~\cite{Schnetter-etal-03b} with the highest resolution
concentrated in the neighbourhood of the individual horizons. For each
of the models studied, we have carried out simulations with fine-grid
resolutions of $h=0.030\,M$ and $h=0.024\,M$, where $M\equiv M_1+M_2$
is the total mass of the system.  For a subset we have carried out a
further evolution at $h=0.018\,M$ to assess the convergence and
estimate the error for the $h=0.024\,M$ results.

The initial data is constructed using the ``puncture''
method~\cite{Brandt97b}, which uses Bowen-York extrinsic curvature and
solves the Hamiltonian constraint equation numerically as
in~\cite{Ansorg:2004ds}. We have considered a sequence of binaries for
which the initial spin of one of the black holes is fixed at
${\boldsymbol S}_2/M^2=0.146{\boldsymbol e}_z$, and, to maximize the
recoil [\textit{cf.} eq.~\eqref{eq:fhh}], the spin of the second black
hole is in the opposite direction and has a modulus which is varied in
steps of $1/4$. As a result, we obtain $5$ sets of binary black holes
having spin ratios $-1,\,-3/4,\ldots,\,0$. The orbital parameters
($m_i$, $x_i$ and $p_i$) for spinning binaries in quasi-circular orbit
are determined using an effective potential method~\cite{Cook94} (a
minimum is found in the binding energy of the system) in such a way
that the black hole masses ($M_i = \sqrt{M_{\rm ir}^2+S_i^2/(4 M_{\rm
ir}^2)}$\,) are equal and $M =1$. The data for the initial separation
and momenta are summarized in Table~\ref{tbl:parameters} and are
chosen so that all of the binaries have the same orbital angular
momentum.

\begin{table}[t]
\begin{ruledtabular}
\begin{tabular}{cccccccc}
&$\pm x/M$&$\pm p/M$&$m_1/M$&$m_2/M$&$S_1/M^2$&$S_2/M^2$ &$M_{_{\rm ADM}}/M$\\
\hline
$r0$ & 3.0205  &  0.1366  &  0.4011 & 0.4009 & -0.1460  & 0.1460  &  0.9856 \\
$r1$ & 3.1264  &  0.1319  &  0.4380 & 0.4016 & -0.1095  & 0.1460  &  0.9855 \\
$r2$ & 3.2198  &  0.1281  &  0.4615 & 0.4022 & -0.0730  & 0.1460  &  0.9856 \\
$r3$ & 3.3190  &  0.1243  &  0.4749 & 0.4028 & -0.0365  & 0.1460  &  0.9857 \\
$r4$ & 3.4100  &  0.1210  &  0.4796 & 0.4034 & -0.0000  & 0.1460  &  0.9859 
\end{tabular}
\end{ruledtabular}
\vskip -0.25cm
\caption{The puncture initial data parameters defining the binaries:
  location ($\pm x/M$), linear momenta ($\pm p/M$), masses ($m_i/M$),
  spins ($S_i/M$) and ADM mass measured at infinity ($M_{_{\rm
      ADM}}$).}
\label{tbl:parameters}
\end{table}

To avoid possible systematic errors and improve the accuracy of the
measurements, the kick velocity has been computed using two different
and independent methods. The first and more traditional one makes use
of the Newman-Penrose quantity $\psi_4$ to provide the rate of change
of the linear momentum in the $i$-direction
as~\cite{Newman80,Campanelli99}
\beq
\label{eq:Pdot}
\frac{dP_i}{dt}=\lim_{r \to \infty} \left\{ \frac{r^2}{16\pi}\int
d\Omega \frac{x_i}{r} \left| \int_{-\infty}^t dt \Psi_4  \right|^2
\right\} \;.
\eeq
The second and novel method, instead, uses a perturbative
wave-extraction procedure~\cite{Rezzolla99a} that calculates
gravitational waves in terms of gauge-invariant, even $Q^{({\rm
    e})}_{\ell m}$ and odd-parity $Q^{({\rm o})}_{\ell m}$ metric
perturbations of the Schwarzschild spacetime after they have been
decomposed into spherical harmonics of spin weight
$-2$~\cite{Nagar05}. Using these quantities it is possible to
reconstruct the momentum fluxes in the $x$ and $y$-directions as
\beq
\label{eq:gi}
\dot{P}_x + {\rm i} \dot{P}_y= \sum^{\infty}_{\ell,m } F(
     Q^{{\rm (e,o)}}_{\ell m},         
     Q^{{\rm (e,o)}\star}_{\ell m},     
\dot Q^{{\rm (e,o)}}_{\ell m},    
\dot Q^{{\rm (e,o)}\star}_{\ell m})\;,
\eeq
where the overdot indicates a (coordinate) time derivative and the
$\star$ complex conjugation [the complete expression for the
  formula~\eqref{eq:gi} will be presented in a longer
  paper~\cite{Koppitz-etal-07b}].
\begin{figure}[tbp!]
\vskip -0.28cm
\centerline{
\resizebox{8.5cm}{!}{\includegraphics[angle=-0]{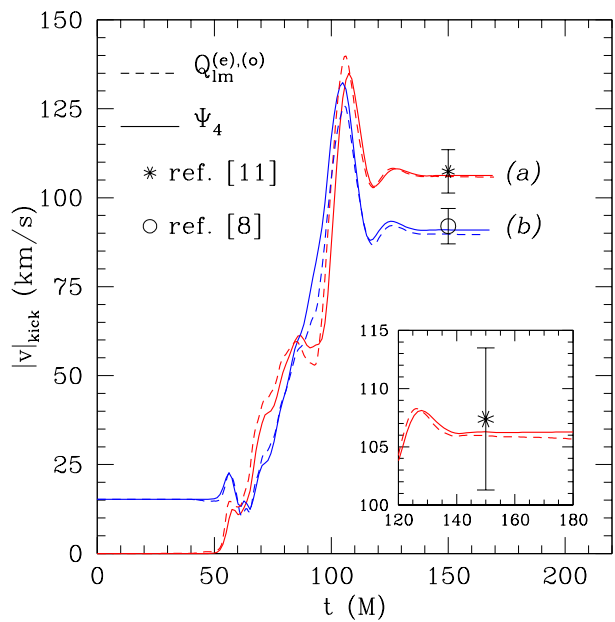}
}}
\caption{Recoil velocity as function of time for a binary system of
  nonspinning black holes with a mass ratio of $2/3$ at an initial
  separation $4.1\,M$. The set of curves \textit{(a)} and \textit{(b)}
  differ in the choice of the integration constant, while the solid
  and dashed lines show the two independent computations of the
  momentum flux [eqs.~\eqref{eq:Pdot} and~\eqref{eq:gi}].}
\label{fig:unequal_mass}
\vskip -0.28cm
\end{figure}

We have validated both methods by measuring the recoil velocity for a
binary system of nonspinning black holes having a mass ratio of
$2/3$ at an initial separation of $4.1\,M$. The results of this
calibration extracted at $r=50\,M$ are shown in
Fig.~\ref{fig:unequal_mass}, which reports the evolution of the kick
velocity using $\Psi_4$ (solid lines), and the gauge-invariant
quantities when the summation in~(\ref{eq:gi}) is truncated to the
first $6$ multipoles (dashed lines), which we have found to be
sufficient to reach convergence. Indicated with symbols are the
estimates and relative error bars obtained by~\cite{Baker:2006nr}
(circle) and by ~\cite{Gonzales06tr} (star).

We note that because the binary system starts evolving at a finite
separation, it will have already gained a net linear momentum which
can influence the value of the final kick. Computing this initial
linear momentum amounts to selecting a proper constant in the
integration of (\ref{eq:Pdot}) or (\ref{eq:gi}). Fortunately, this is
rather straightforward to do and amounts to determining the direction
in 3-space in which the center of mass of the system is moving
initially. In practice, we plot the evolution of the $x$ and
$y$-components of the kick velocity (the $z$-component is zero because
of symmetry) and calculate the vector to the centre of the spiral
generated as the evolution proceeds. This vector is then composed with
the final one, yielding the final kick; note that being a vector this
integration constant \textit{is not} simply an additive constant for
the kick velocity $|v|_{\rm kick}$. In Fig.~\ref{fig:unequal_mass}, we
have plotted the effect of including this constant, comparing the case
where it is set to zero [set of curves \textit{(a)}], with a value set
by extrapolating the recoil backwards to compensate for the small but
nonzero initial linear momentum [set of curves \textit{(b)}]. In the
first case we find agreement with~\cite{Gonzales06tr}, while in the
second case the good agreement is with~\cite{Baker:2006nr}.

\begin{figure}[tbp!]
\vskip -0.28cm
\centerline{
\resizebox{8.5cm}{!}{\includegraphics[angle=-0]{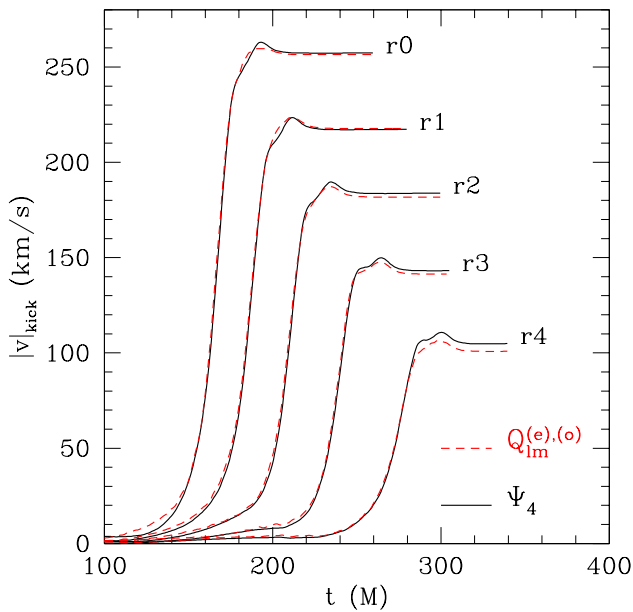}
}}
\caption{Recoil velocity as function of time for the sequence of runs
  \textit{i.e.,} from $r0$ with $-a_1 = a_2 = 0.586$, to r4 with $a_1 =
  0, a_2=0.586$). Note that the merger is delayed for smaller values
  of $|a_1|$.}
\label{fig:vkick_vs_time}
\vskip -0.28cm
\end{figure}

A validation of this procedure is also rather straightforward: only an
accurate estimate of the initial momentum yields a monotonic evolution
of the kick velocity (or, in the case of very close binaries, reduces
the oscillations considerably); any different choice would yield the
oscillations seen in curves \textit{(a)} (\textit{cf.,} Fig. 1 of
ref.~\cite{Baker:2006nr} or Fig. 3 of
ref.~\cite{Gonzales06tr}). Clearly, selecting the correct integration
constant becomes less important as the separation in the binary is
increased (see also the discussion below), but it can easily lead to
errors of $10\%$ or more for the rather close binaries considered
here. A more detailed discussion on the integration constant will be
given in~\cite{Koppitz-etal-07b}.


\label{se:results}

\noindent\emph{Results.~} In Fig.~\ref{fig:vkick_vs_time} we show the
evolution of the recoil velocity for the $5$ binaries considered, with
the dashed lines referring to the gauge-invariant extraction and the
solid lines to the one using $\Psi_4$; both quantities are extracted
at $50\,M$ which we have found to be the smallest needed distance for
consistent results.
\begin{figure}[tbp!]
\vskip -0.28cm \centerline{
\resizebox{7.5cm}{!}{\includegraphics[angle=-0,height=65mm]{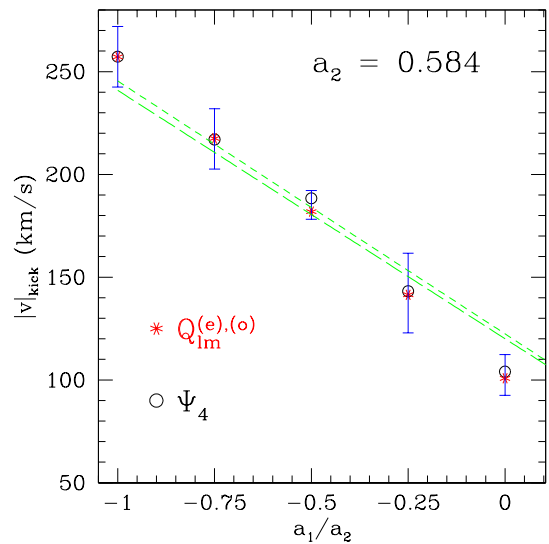}
}}
\caption{Kick velocities and error bars for different spin ratios; the
  dashed lines show a linear fit of all the data when the point at
  $a_1/a_2=1$ is given an infinite weight since $|v|_{\rm kick} = 0$
  for $a_1=a_2$.}
\label{fig:final_values}
\vskip -0.28cm
\end{figure}

A number of interesting features are worth remarking. Firstly, all of
the curves show a monotonic growth as a result of a suitable choice
for the integration constant. Secondly, the two types of measurements
agree to very good precision. Thirdly, binaries that have a spin ratio
closer to zero merge progressively later. In the case of the binaries
considered here this effect is partially masked by the fact that the
binary $r0$ is at a smaller separation than $r4$ (\textit{cf.,}
Table~\ref{tbl:parameters}). Bearing this in mind, it is however
apparent that the growth rate of the kick velocity (and hence the
rapidity of the inspiral), increases with the asymmetry in the
spins. Fourthly, increasing the initial separation for a binary with
$a_1/a_2=-1$ does not change significantly the integration constant
chosen for $r0$, thus indicating that the kick estimate for the latter
is robust. Finally, as in unequal-mass binaries, the largest
contribution to the kick comes from the final parts of the inspiral
and is dominated by the last orbit. However, unlike equal-mass
binaries, the post-merger evolution of the kick velocity is not
modified substantially by the quasi-normal mode ringing (\textit{cf.,}
Figs.~\ref{fig:vkick_vs_time} and~\ref{fig:unequal_mass}), with the
final kick velocity being only slightly smaller than the maximum one
reached during the evolution.

As predicted by the PN expression (\ref{eq:fhh}), the final velocities
shown in Fig.~\ref{fig:vkick_vs_time} exhibit a linear dependence with
the spin ratio, and this is shown in Fig.~\ref{fig:final_values},
which reports the asymptotic kick velocities when measured with
$\Psi_4$ (open circles) or with the gauge-invariant perturbations
(stars). Also indicated are the error bars which include errors from
the determination of the integration constants, from the dependence of
the waveforms on the extraction radii, and from the truncation error.

The data points in Fig.~\ref{fig:final_values} are not the only ones
available and indeed a binary system with $a_1/a_2=1$ is bound to
produce a zero kick velocity.  The dashed lines in
Fig.~\ref{fig:final_values} represent a linear fit of all the data
when the point at $a_1/a_2=1$ is given an infinite weight to account
that $|v|_{\rm kick} = 0$ when $a_1=a_2$ (short-dashed line for
$\Psi_4$ and long-dashed for $Q^{\rm (e,o)}_{\ell m}$). These lines
are only indicative and bear a physical significance only if the
linear dependence should hold for all the possible values of the
spin-ratio.


\noindent\emph{Conclusions~} We have calculated the recoil velocity
from the inspiral and merger of binary black-hole systems with equal
masses but unequal spins. To increase the accuracy and remove
systematic errors, we have performed our measurements using both the
Newman-Penrose quantity $\Psi_4$, as well as gauge-invariant metric
perturbations of a Schwarzschild spacetime. The two methods agree to
very good precision and indicate that the recoil velocity produced by
spin-asymmetries can be considerably larger than that expected from
asymmetries in the masses. More specifically, we have found a maximum
recoil velocity of $257 \pm 15$ km/s for a system having a spin ratio
$a_1/a_2 = -1$ and $a_2=0.584$. In addition, a linear scaling holds
between the kick velocity and the spin ratio, thus qualitatively
confirming the PN expectation. A linear extrapolation to the case of
extremal black holes, \textit{i.e.,} $a_1=1=-a_2$, suggests a maximal
value for the kick velocity from unequal-spin, but equal-mass binaries
of $\simeq 440$ km/s. Such a recoil velocity would be more than twice
that produced by non-spinning but unequal-mass binaries.

At the time of the submission of this paper a number of complementary
studies have also been submitted~\cite{Herrmann:2007ac,
Campanelli:2007cg,Gonzalez:2007hi}, which support the results found
here as well as examine non-aligned spin configurations.


\noindent\emph{Acknowledgments.~} It is a pleasure to thank Alessandro
Nagar Bernard Schutz, Ed Seidel and Ryoji Takahashi for useful
discussions. The computations were performed at AEI, LSU/LONI, NERSC,
and at NCSA (grant MCA02N014). This work was also supported by the DFG
(grant SFB TR/7).


\bibliographystyle{AEIBibtex/apsrev-nourl}
\bibliography{AEIBibtex/aeireferences}

\end{document}